\documentclass[showpacs, superscriptaddress, showkeys, nofootinbib, aps, prd, preprintnumbers, twocolumn]{revtex4-1}

\usepackage{subfigure}
\usepackage{graphicx}

\usepackage{bm}

\usepackage{amsmath, amssymb}
\usepackage{times}

\usepackage[usenames]{xcolor}
\definecolor{mpl_red}{HTML}{D62728}

\usepackage[breaklinks, plainpages=false, colorlinks=true, anchorcolor=blue!50!black, citecolor=blue!50!black, linkcolor=blue!50!black, urlcolor=mpl_red, bookmarks=false]{hyperref}
\usepackage{natbib}

\def\aj{\rm{AJ}}
\def\apj{\rm{ApJ}}
\def\apjl{\rm{ApJL}}
\def\apjs{\rm{ApJS}}

\def\aaps{\rm{A\&AS}}
\def\mnras{\rm{MNRAS}}

\def\nat{\rm{Nature}}

\def\araa{\rm{ARA\&A}}
\def\aapr{\rm{A\&A~Rev.}}

\def\ssr{\rm{Space Sci. Rev.}}

\def\pasa{\rm{PASA}}

\begin{document}

\title{Observations of Radio Magnetars with the Deep Space Network}

\newcommand{\CaltechPhysics}{Division of Physics, Mathematics, and Astronomy, California Institute of Technology, Pasadena, CA 91125, USA}
\newcommand{\JPL}{Jet Propulsion Laboratory, California Institute of Technology, Pasadena, CA 91109, USA}
\newcommand{\NDSEG}{$^{\text{3}}$~NDSEG Research Fellow.}
\newcommand{\NSF}{$^{\text{4}}$~NSF Graduate Research Fellow.}

\author{Aaron~B.~Pearlman}
\email{aaron.b.pearlman@caltech.edu}
\thanks{NDSEG Research Fellow; NSF Graduate Research Fellow.}
\affiliation{\CaltechPhysics}

\author{Walid~A.~Majid}
\affiliation{\JPL}
\affiliation{\CaltechPhysics}

\author{Thomas~A.~Prince}
\affiliation{\CaltechPhysics}
\affiliation{\JPL}

\newcommand{\PreprintSpacing}{~~~~~~~~~~~~~~~~~~~~~~~~~~~~~~~~~~~~~~~~~~~~~~~~~~~~~~~~~~~~~~~~~~~~~~~~~~~~~~~~~~~~~~~~~~~~~~~~~~~~~~~~~~~~~~~~~~~~~~~~~~~~~~~~~~~~~~~~~~~~~~~~~~~~~~~~~~~~~~~~~~~~~~}
\preprint{Accepted for publication in Advances in Astronomy on 2019~January~27\PreprintSpacing}

\begin{abstract}

The Deep Space Network~(DSN) is a worldwide array of radio telescopes that supports NASA's interplanetary spacecraft missions. When the DSN~antennas are not communicating with spacecraft, they provide a valuable resource for performing observations of radio magnetars, searches for new pulsars at the Galactic Center, and additional \text{pulsar-related} studies. We describe the DSN's capabilities for carrying out these types of observations. We also present results from observations of three radio magnetars, PSR~J1745--2900, PSR~J1622--4950, and XTE~J1810--197, and the transitional magnetar candidate, PSR~J1119--6127, using the DSN~radio telescopes near Canberra, Australia.

\end{abstract}

\pacs{}
\keywords{}

\maketitle

\section{Introduction}
\label{Section:Introduction}

\setcounter{footnote}{2}

\begin{figure*}[t]
	\centering
	\includegraphics[trim=0cm 0cm 0cm 0cm, clip=false, scale=0.45, angle=0]{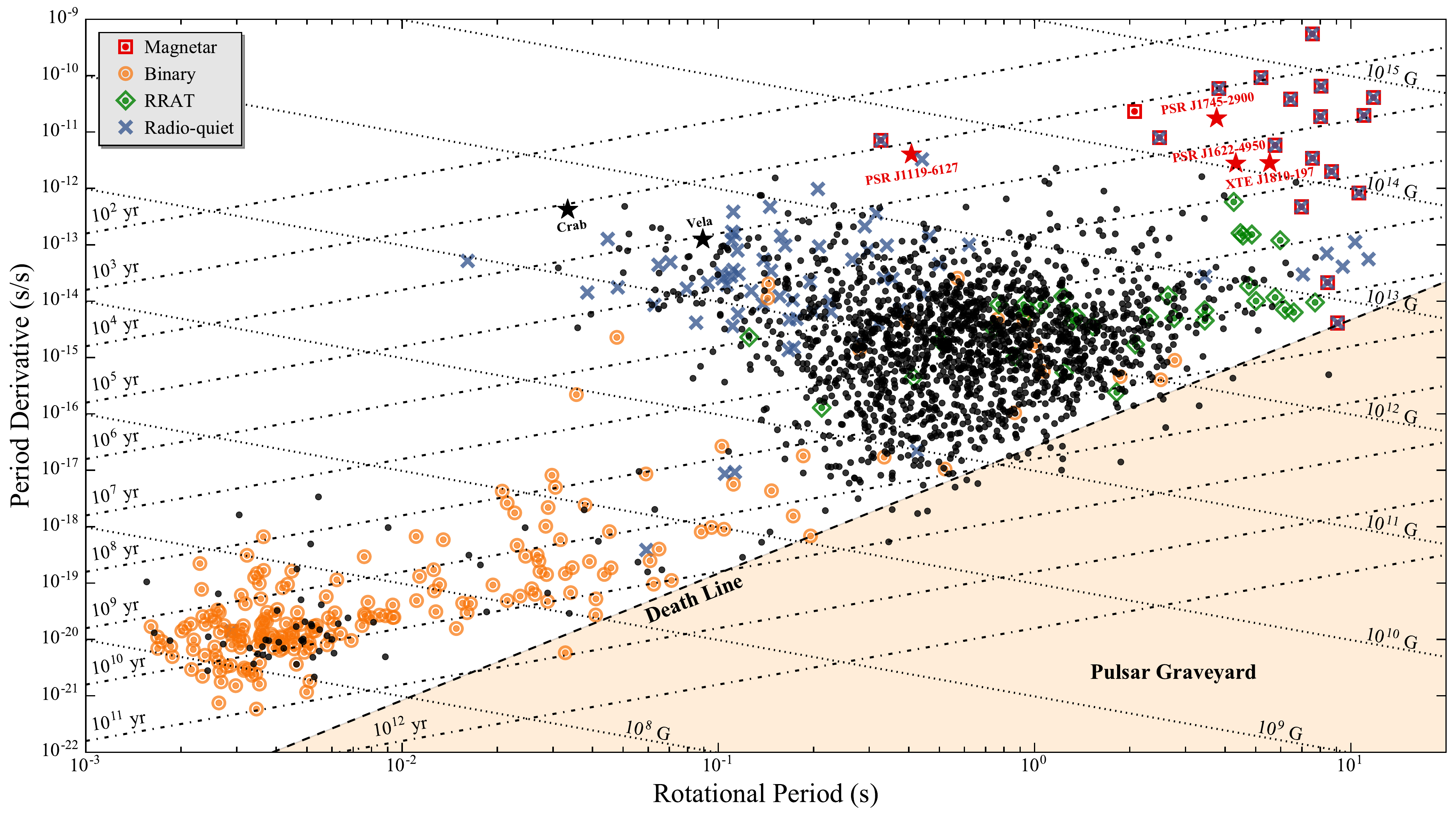}
	\caption{$P$--$\dot{P}$ diagram of pulsars in the Australia Telescope National Facility~(ATNF) pulsar catalog\protect\footnotemark[1]~\citep{Manchester+2005}. The four radio magnetars, PSR~J1745--2900, PSR~J1119--6127, PSR~J1622--4950, and XTE~J1810--197, discussed in this paper are labeled using red stars. Magnetars~(red squares), rotating radio transients~(RRATs; green diamonds), and millisecond pulsars~(MSPs; orange circles) are also shown, along with the population of rotation-powered radio pulsars~(black circles). Radio-quiet pulsars are indicated with blue crosses and include many of the magnetars shown in red. Lines of constant magnetic field and characteristic age are derived assuming a constant braking index of $n$\,$=$\,3. The radio pulsar death line is given by the model in Equation~(4) of~\citet{Zhang+2000}.}
	\label{Figure:Figure1}
	
	\footnotetext{See https://www.atnf.csiro.au/people/pulsar/psrcat.}
\end{figure*}

Magnetars are young neutron stars with very strong magnetic fields~($B$\,$\approx$\,10$^{\text{13}}$--10$^{\text{15}}$\,G). They have rotational periods between $\sim$2--12\,s and larger than average spin-down rates compared to other pulsars, placing them in the top right region of the $P$--$\dot{P}$ diagram~(see~Figure~\ref{Figure:Figure1}). Magnetars are primarily powered by the decay of their enormous magnetic fields, which serves as an energy source for their transient emission behavior~\citep{Duncan+1992, Thompson+1995, Thompson+1996}. It is thought that magnetars comprise at least 10\% of the young neutron star population~\citep{Kaspi+2017}, and they tend to be more concentrated toward the inner part of the Galaxy~\citep{Olausen+2014}.

\begin{figure*}[t]
	\centering
	\includegraphics[trim=0cm 0cm 0cm 0cm, clip=false, scale=0.8, angle=0]{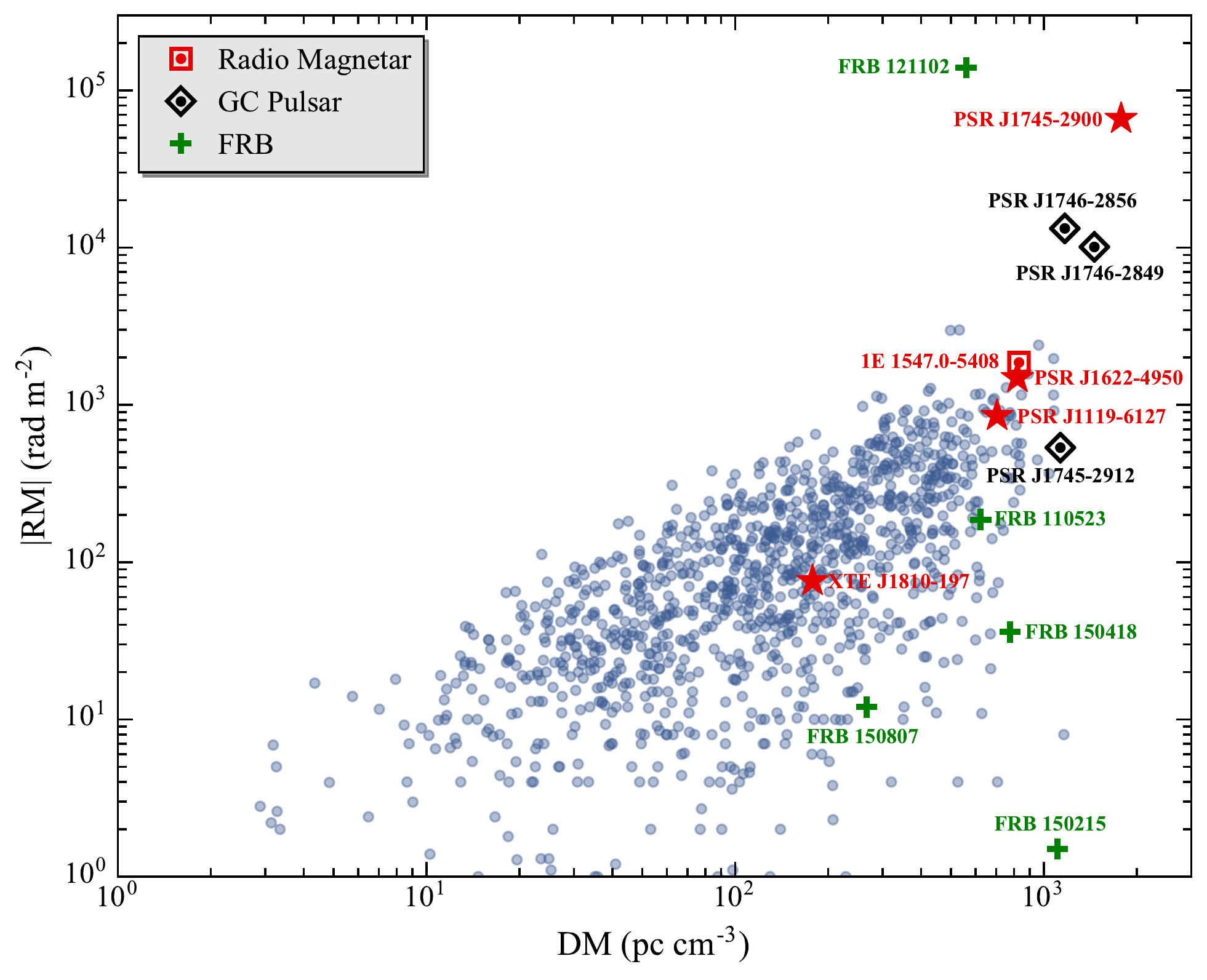}
	\caption{Magnitude of the Faraday rotation measure~(RM) versus dispersion measure~(DM) for all known pulsars~(blue circles) in the Australia Telescope National Facility~(ATNF) pulsar catalog\protect\footnotemark[1]~\citep{Manchester+2005}. We label the four radio magnetars considered in this paper, PSR~J1745--2900, PSR~J1119--6127, PSR~J1622--4950, and XTE~J1810--197, using red stars. The one other radio magnetar, 1E~1547.0--5408, is indicated by a red square. We also show the Galactic Center~(GC) pulsars using black diamonds and the four fast radio bursts~(FRBs) with RM~measurements in the FRB catalog\protect\footnotemark[2]~\citep{Petroff+2016} using green crosses.}
	\label{Figure:Figure2}
	
	\footnotetext{See https://www.atnf.csiro.au/people/pulsar/psrcat.}
	\footnotetext{See http://frbcat.org.}
\end{figure*}

There are currently 29~known magnetars and 2~additional magnetar candidates (normally rotation-powered pulsars), the latter exhibiting episodes of magnetar-like behavior. More than $\sim$2600 pulsars have been discovered, but only four of these are radio magnetars: PSR~J1745--2900, PSR~J1622--4950, XTE~J1810--197, and 1E~1547.0--5408. Thus, radio magnetars are exceptionally rare and constitute $\lesssim$\,0.2\% of the pulsar population. They also have large dispersion measure~(DM) and Faraday rotation measure~(RM) values compared to ordinary radio pulsars, which suggest that they inhabit extreme magneto-ionic environments~(see~Figure~\ref{Figure:Figure2}). A detailed list of properties associated with known magnetars can be found in the McGill Magnetar Catalog\footnote{See http://www.physics.mcgill.ca/$\sim$pulsar/magnetar/main.html.}~\citep{Olausen+2014}.

Radio magnetars often have flat or inverted radio spectra, and their radio emission is highly linearly polarized~(e.g.,~\citep{Camilo+2006, Camilo+2007c, Levin+2010, Keith+2011, Levin+2012}). As a result, they are capable of being detected at very high radio frequencies~(e.g.,~\citep{Camilo+2007c, Torne+2015, Torne+2017}). The flux densities, spectral indices, and pulse shapes of radio magnetars can also change on short timescales~(e.g.,~\citep{Camilo+2007a, Lazaridis+2008, Levin+2010, Levin+2012, Scholz+2017, Pearlman+2016, Pearlman+2017, Pearlman+2018}), and their radio pulse profiles often display multiple emission components, which can vary significantly across multiple radio frequencies~(e.g.,~\citep{Camilo+2006, Camilo+2007a, Camilo+2007d, Camilo+2008, Camilo+2016, Pearlman+2018}). Individual pulses from radio magnetars are typically comprised of narrow sub-pulses, which can be exceptionally bright. However, they are unlike the giant pulses emitted by the Crab pulsar~\citep{Cordes+2004, Camilo+2006, Pearlman+2018}. The morphology of these pulses can also change substantially between rotations~(e.g.,~\citep{Camilo+2006, Pearlman+2018}). Irregular timing behavior, including glitches~(sudden increases in the pulsar's rotation frequency), is also commonly observed from radio magnetars~(e.g.,~\citep{Camilo+2007a, Camilo+2008, Weltevrede+2011, Antonopoulou+2015, Archibald+2016a, Camilo+2016, Archibald+2018}), and their radio emission has been reported to episodically disappear and suddenly reactivate~\citep{Burgay+2009, Camilo+2009, Burgay+2016a, Burgay+2016b, Camilo+2016, Majid+2016, Majid+2017}. Radio active and quiescent magnetars can also emit short \text{X-ray} bursts~\citep{Degenaar+2013, Kennea+2013, Archibald+2016a, Gogus+2016}. The assortment of behavior listed here points to an underlying connection between high magnetic field radio pulsars and magnetars~\citep{Camilo+2006, Archibald+2016a, Majid+2016, Pearlman+2018}.

In this paper, we discuss recent observations of three radio magnetars, PSR~J1745--2900, PSR~J1622--4950, XTE~J1810--197, and the transitional magnetar candidate, PSR~J1119--6127, using the Deep Space Network~(DSN) radio telescopes. This paper is not intended to be a comprehensive review of the vast literature on radio magnetars. Instead, we focus on recent observational results on these particular radio magnetars using the DSN antennas. For a more complete review of magnetars, we refer the interested reader to some of the available review articles on the subject~(e.g.,~\citep{Woods+2006, Kaspi2010, Rea+2011, Mereghetti+2015, Turolla+2015, Kaspi+2016, Kaspi+2017}). In Section~\ref{Section:DSN}, we describe the DSN radio dishes and the system's observing capabilities. We discuss our observational results on each of the four magnetars listed above in Sections~\ref{Section:GC_magnetar}--\ref{Section:XTE_J1810_197}. A summary is provided in Section~\ref{Section:Conclusions}. This paper was prepared in response to an invited solicitation for a dedicated special issue on magnetars.



\section{The Deep Space Network}
\label{Section:DSN}

The~DSN consists of an array of radio telescopes at three locations (Goldstone, California; Madrid, Spain; and Canberra, Australia). Each of these sites is approximately equally separated in terrestrial longitude and situated in a relatively remote location to shield against radio-frequency interference~(RFI). With multiple radio antennas at each site, the~DSN covers both celestial hemispheres and serves as the spacecraft tracking and communication infrastructure for NASA's deep space missions. The three DSN complexes each include a 70\,m diameter antenna, with a surface suitable for radio observations at frequencies up to 27\,GHz. In addition, each site hosts a number of smaller 34\,m diameter radio telescopes, which are capable of observations as high as 32\,GHz. Each antenna is equipped with multiple high efficiency feeds, highly sensitive cryogenically cooled receivers, and dual (circular) polarization capabilities. When the DSN~antennas are not communicating with spacecraft, they may be used for radio astronomy and other radio science applications.

Recently, all three sites have been upgraded with \text{state-of-the-art} pulsar processing backends that enable data recording with high time and frequency resolution. The~DSN telescopes are able to perform radio observations at the following standard frequency bands: $L$-band (centered at 1.5\,GHz), $S$-band (centered at 2.3\,GHz), $X$-band (centered at 8.4\,GHz), and $Ka$-band (centered at 32\,GHz). In addition, the 70\,m radio dish in Canberra~(see Figure~\ref{Figure:Figure4}) is outfitted with a dual beam $K$-band feed covering 17--27\,GHz. These capabilities are currently being used in various \text{pulsar-related} programs, which include high frequency, ultra-wide bandwidth searches for pulsars in the Galactic Center~(GC), high frequency monitoring of radio magnetars~\citep{Majid+2016, Pearlman+2016, Majid+2017, Pearlman+2017, Pearlman+2018, Majid+2019, Pearlman+2019a, Pearlman+2019b}, multifrequency studies of giant pulses from the Crab pulsar~\citep{Majid+2011, Majid+2013}, and high frequency searches for fast radio bursts~(FRBs).

The~DSN radio telescopes are particularly \text{well-suited} for monitoring radio magnetars. These instruments allow for high cadence observations, which are important for tracking changes in the flux densities, pulse profile shapes, spectral indices, and single pulse behavior of radio magnetars, all of which can vary on daily timescales. High frequency observations are also essential because the spectral indices of radio magnetars are quite flat or inverted on average. In fact, the GC~magnetar, PSR~J1745--2900 (see Section~\ref{Section:GC_magnetar}), has been detected at record high radio frequencies~\citep{Torne+2015, Torne+2017}. The~DSN antennas are also capable of providing simultaneous, dual band observations with both circular polarizations, which is essential for accurate spectral index measurements and polarimetric studies. Additionally, since the large 70\,m dishes have very low system temperatures, they are ideal for studying the morphology of single pulses from radio magnetars. We refer the reader to several of our recent papers~(see~\citep{Majid+2011, Majid+2013, Majid+2016, Pearlman+2016, Majid+2017, Pearlman+2017, Pearlman+2018, Majid+2019, Pearlman+2019a, Pearlman+2019b}), which are summarized in Sections~\ref{Section:GC_magnetar}--\ref{Section:XTE_J1810_197}.



\begin{figure*}[t]
	\centering
	\includegraphics[trim=0cm 0cm 0cm 0cm, clip=false, scale=0.6, angle=0]{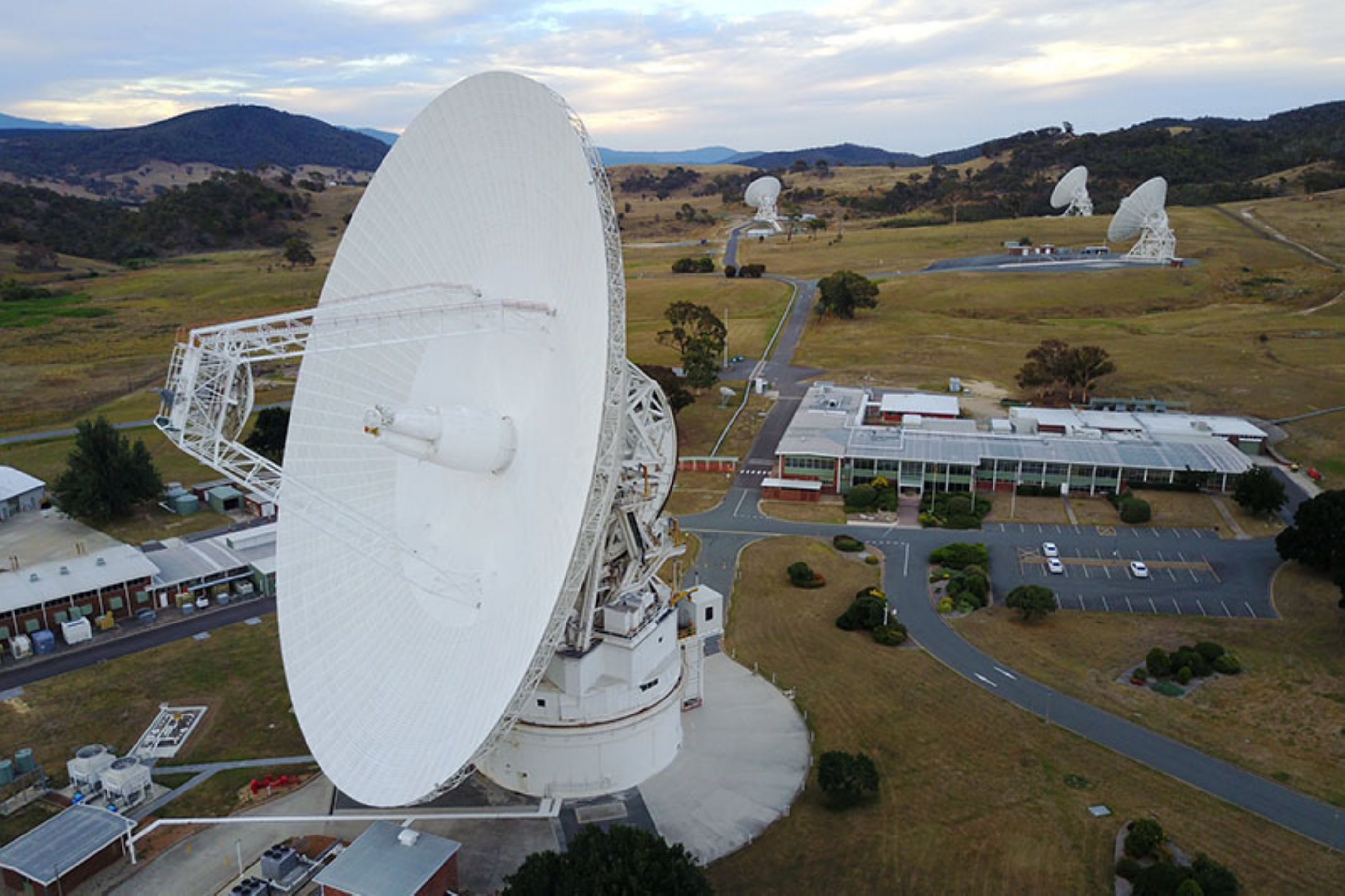}
	\caption{Image of the Canberra Deep Space Communication Complex~(CDSCC). The 70\,m telescope~(DSS-43) is shown in the foreground and the three 34\,m beam waveguide antennas (DSS-34, DSS-35, and DSS-36) are shown in the background. Image credit: CDSCC\protect\footnotemark[1].}
	\label{Figure:Figure3}
	
	\footnotetext{See https://www.cdscc.nasa.gov/Pages/antennas.html.}
\end{figure*}


\section{PSR~J1745--2900: The Galactic Center Magnetar}
\label{Section:GC_magnetar}

The GC~magnetar, PSR~J1745--2900, was serendiptiously discovered by the Neil Gehrels \textit{Swift} Observatory after a hard \text{X-ray} burst was detected on 2013~April~24~\citep{Degenaar+2013, Kennea+2013}. The magnetar has DM~(1778\,$\pm$\,3\,pc\,cm$^{\text{--3}}$) and RM~(--66960\,$\pm$\,50\,rad\,m$^{\text{--3}}$) values that are larger in magnitude than any known pulsar~\citep{Eatough+2013a}~(see~Figure~\ref{Figure:Figure3}). It is located $\sim$0.1\,pc from the Galaxy's central 4\,$\times$\,10$^{\text{6}}$\,$M_{\odot}$ black hole, Sagittarius~A$^{\ast}$~(Sgr~A$^{\ast}$)~\citep{Bower+2015}, making it an excellent probe of the magneto-ionic environment near the inner region of the Galaxy. We recently carried out simultaneous radio observations of PSR~J1745--2900 at 2.3~and~8.4\,GHz during four separate epochs between 2015~July 30 and 2016~August~20 using the 70\,m~DSN~radio telescope,~DSS-43~(see~Section~\ref{Section:DSN})~\citep{Pearlman+2018}. The observational parameters used for this study are provided in~\citet{Pearlman+2018}. Here, we discuss our measurements of the magnetar's radio profile shape, flux density, radio spectrum, and single pulse behavior, which are also described in detail in~\citet{Pearlman+2018}.

\begin{figure*}[t]
	\centering
	\includegraphics[trim=0cm 0cm 0cm 0cm, clip=false, scale=0.455, angle=0]{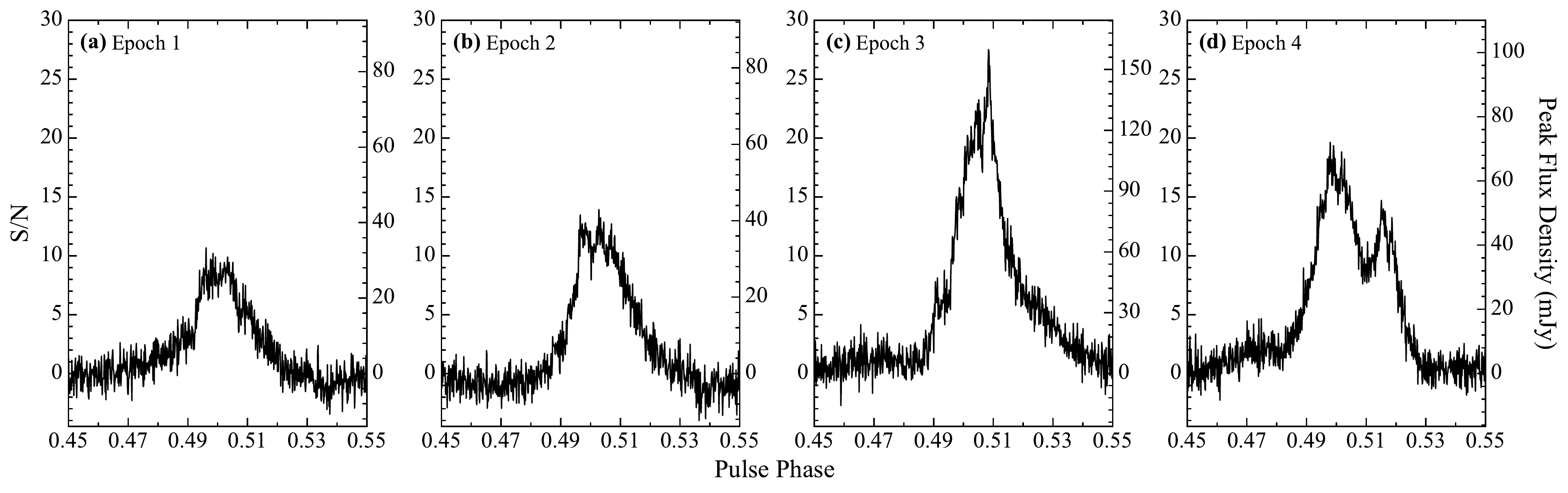}
	\caption{Average $X$-band pulse profiles of the GC~magnetar from observations with the 70\,m~DSN~radio telescope,~DSS-43, carried out on (a)~2015~July~30, (b)~2015~August~15, (c)~2016~April~1, and (d)~2016~August~20. This figure was adapted from~\citet{Pearlman+2018}.}
	\label{Figure:Figure4}
\end{figure*}

Radio pulsations were detected at a period of $P$\,$\approx$\,3.77\,s in all of our observations~\citep{Pearlman+2018}. The average $X$-band pulse profiles, shown in Figure~\ref{Figure:Figure4}, are single-peaked during epochs~1--3 and double-peaked during epoch~4. The $S$-band pulse profiles are not shown here since the pulsed emission was significantly weaker at this frequency. The mean flux density at $S$-band was noticeably variable~\citep{Pearlman+2018}, and the $X$-band flux densities on 2015~July~30 and~August~15 were smaller by a factor of~$\sim$7.5 compared to measurements performed $\sim$5~months earlier by~\citet{Torne+2017}. Our measurements on 2016~April~1 and August~20 indicated that the $X$-band flux density more than doubled since 2015~August~15~\citep{Pearlman+2018}.

Multifrequency radio observations of PSR~J1745--2900 revealed that its radio spectrum is often relatively flat or inverted~\citep{Eatough+2013a, Torne+2015, Torne+2017}, which is typical of most radio magnetars. However, its radio spectrum can also significantly steepen to levels comparable to ordinary radio pulsars~\citep{Pennucci+2015, Pearlman+2018}, which have an average spectral index of \mbox{$\langle\alpha\rangle$\,$=$\,--1.8\,$\pm$\,0.2}~\citep{Maron+2000}. During epochs~1--3, \citet{Pearlman+2018} found that the magnetar exhibited a significantly negative average spectral index of $\langle\alpha\rangle$\,$=$\,--1.86\,$\pm$\,0.02 when the average pulse profile was single-peaked, which is comparable to the steep spectrum derived by~\citet{Pennucci+2015} between 2~and~9\,GHz. The spectral index then significantly flattened to $\alpha$\,$>$\,--1.12 during epoch~4 when the profile displayed an additional component~\citep{Pearlman+2018}.

\begin{figure*}[b!]
	\centering
	\includegraphics[trim=0cm 0cm 0cm 0cm, clip=false, scale=0.44, angle=0]{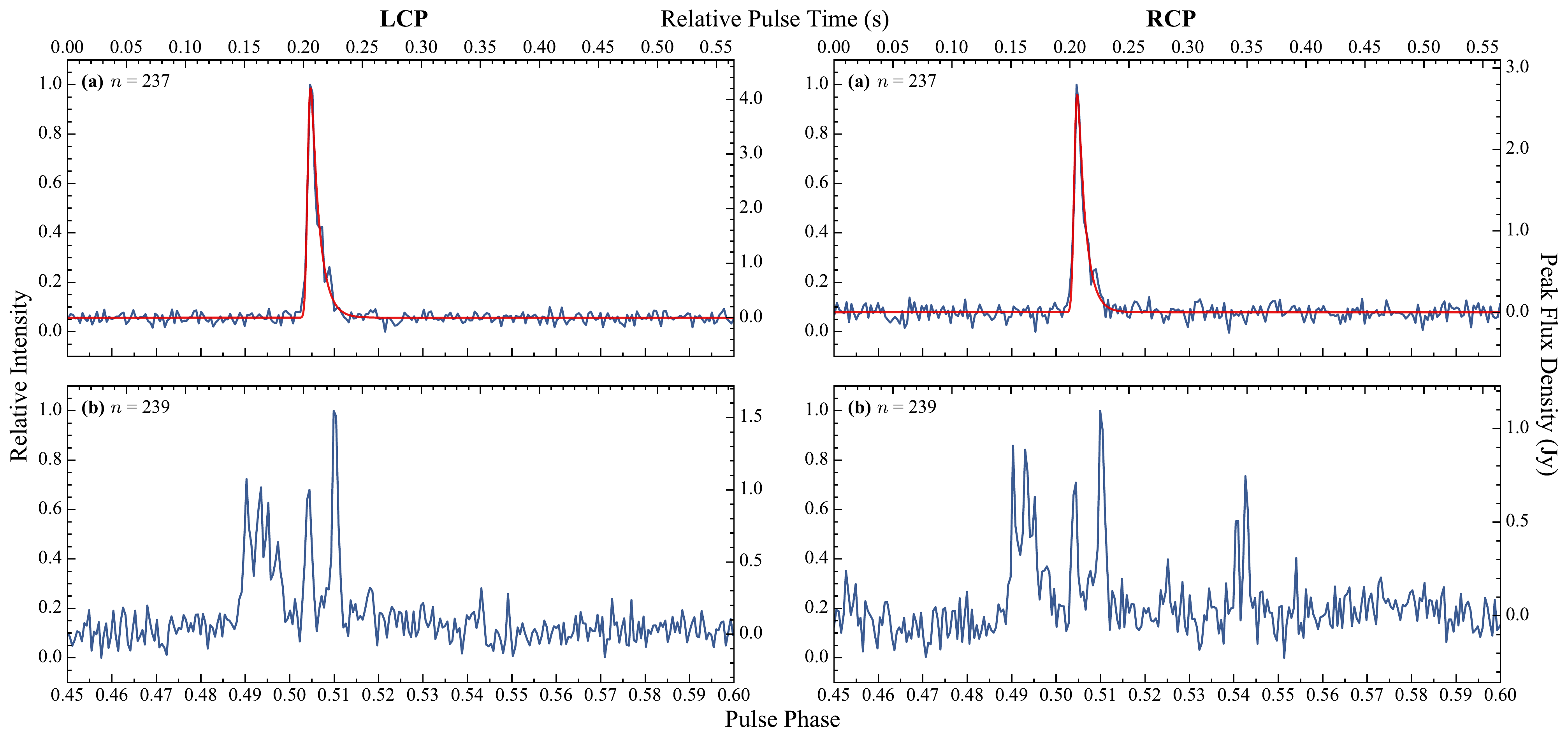}
	\caption{Examples of bright single pulses detected from the GC~magnetar at 8.4\,GHz during epoch~3~(2016~April~1) with the 70\,m~DSN~radio telescope,~DSS-43. We show the single pulse profiles measured in the left circular polarization~(LCP) and right circular polarization~(RCP) channels during pulse cycles (a)~$n$\,$=$\,237 and (b)~$n$\,$=$\,239, where pulse numbers are referenced with respect to the start of the observation. The best-fit thin scattering screen model, described in the Appendix of~\citet{Pearlman+2018}, is overlaid in red on the single pulses in the top row. A pulse broadening timescale of $\tau_{d}^{\text{LCP}}$\,$=$\,7.1\,$\pm$\,0.2\,ms and $\tau_{d}^{\text{LCP}}$\,$=$\,6.7\,$\pm$\,0.3\,ms was measured in the corresponding polarization channels~\citep{Pearlman+2018}. This figure was adapted from~\citet{Pearlman+2018}.}
	\label{Figure:Figure5}
\end{figure*}

\citet{Pearlman+2018} also performed an analysis of single pulses detected at 8.4\,GHz during epoch~3, which displayed the brightest pulses. They found that the single pulse structure observed from the GC~magnetar was extremely variable in time, and the pulse morphology can be entirely different between successive rotations~\citep{Pearlman+2018}~(e.g.,~see~Figure~\ref{Figure:Figure5}). Giant pulses, with flux densities more than ten times the mean flux level, and pulses with multiple emission components were detected during many of the magnetar's rotations~\citep{Pearlman+2018}. These giant pulses are different in nature from those emitted by the Crab pulsar~\citep{Cordes+2004}. There is also some evidence that the brightest emission component appears first during a given rotation and may trigger additional weaker outbursts~\citep{Pearlman+2018}.

The observed flux density distribution of the single pulses could not be described by a log-normal distribution due to these giant pulses, and bright single pulses can sometimes form a high flux tail~\citep{Lynch+2015, Pearlman+2018}. This is similar to the single pulse behavior reported from the transient radio magnetar, XTE~J1810--197, whose pulse-energy distribution is well described by a log-normal distribution at lower energies and a power-law at higher energies~\citep{Serylak+2009}. In particular, \citet{Pearlman+2018} measured a scaling exponent of $\Gamma$\,$=$\,--7\,$\pm$\,1 from a power-law fit to the distribution of single pulse flux densities with peak fluxes greater than 15~times the mean level. The pulse intensity distribution is likely variable in time since a high flux tail has not persistently been observed from the GC~magnetar~\citep{Lynch+2015, Yan+2015, Gelfand+2017, Pearlman+2018}. No correlation has been found between the peak flux density and the number of emission components in the single pulses~\citep{Pearlman+2018}. In addition, an earlier single pulse analysis by~\citet{Yan+2015} at 8.6\,GHz revealed no obvious correlation between the width and the peak flux density of the GC magnetar's strongest pulses, and there was no evidence indicating sub-pulse drifting in their observations.

\citet{Pearlman+2018} found that the typical intrinsic pulse width of the emission components was $\sim$1.8\,ms, and they reported a prevailing delay time of $\sim$7.7\,ms between successive components. Additionally, their analysis showed that some of the emission components at later pulse phases were detected more strongly in one of the circular polarization channels, which suggests that some of the magnetar's emission components may be more polarized than others~(e.g.,~see~Figure~\ref{Figure:Figure5}(b))~\citep{Pearlman+2018}. The overall single pulse behavior during epoch~3 can be explained by fan beam emission with a width of $\pm$7$^{\circ}$, and tapering of the fan beam may give rise to fainter emission components at later pulse phases~\citep{Pearlman+2018}.

The GC~magnetar's emission region is thought to emit pulses fairly regularly since bright single pulses were detected during almost all rotations. During epoch~3, bright pulses were detected during $\sim$70\% of the GC~magnetar's rotations, but often not at precisely the same phase~\citep{Pearlman+2018}. At higher radio frequencies~($\sim$45\,GHz), \citet{Gelfand+2017} found that bright pulses, with an average width of $\sim$4.6\,ms, were produced during a similar fraction of the magnetar's rotations. However, radio observations at 3.1\,GHz have shown that the GC~magnetar can exhibit brief periods of pulsar nulling~\citep{Yan+2018}. Strong single pulses have also been detected at radio frequencies up to 154\,GHz~\citep{Torne+2015}, which suggests a broadband underlying emission mechanism.

\citet{Pearlman+2018} discovered significant frequency structure over bandwidths of $\sim$100\,MHz in many of the single pulse emission components, which is the first time such behavior has been observed from a radio magnetar. They argued that these features could be produced by strong lensing from refractive plasma structures located near the inner part of the Galaxy, but may also be intrinsic to the magnetar and possibly similar to the banded structures observed in the Crab pulsar's High-Frequency Interpulse~\citep{Hankins+2016}. Diffractive interstellar scintillation and instrumental effects were both ruled out as possible origins of this behavior~\citep{Pearlman+2018}. If these features are indeed due to interstellar plasma lensing, then this suggests that a magneto-ionic medium close to the~GC can boost the observed flux densities of pulses by more than an order of magnitude~\citep{Pearlman+2018}. This behavior is reminiscent of the structure observed in pulses from the repeating FRB~121102~\citep{Spitler+2016, Gajjar+2018, Michilli+2018} and may indicate a connection with the larger population of~FRBs. Furthermore, the GC~magnetar and the repeating FRB~121102 have similar~DM and RM~values~(see Figure~\ref{Figure:Figure2}) and emit pulses with similar morphology. An extragalactic magnetar near a massive black hole, perhaps not unlike PSR~J1745--2900, is one of the currently favored progenitor theories for FRBs~(e.g.,~\citep{Pen+2015, Metzger+2017, Michilli+2018}).

Recently, \citet{Pearlman+2018} showed that the emission components comprising the GC magnetar's single pulses can be substantially broadened~(e.g.,~see~Figure~\ref{Figure:Figure5}(a)). A characteristic single pulse broadening timescale of $\langle\tau_{d}\rangle$\,$=$\,6.9\,$\pm$\,0.2\,ms was reported at 8.4\,GHz~\citep{Pearlman+2018}. The pulse broadening magnitude was also found to be variable between pulses detected during consecutive pulse cycles and between pulse components in the same pulsar rotation~\citep{Pearlman+2018}. \citet{Pearlman+2018} argued that this behavior could be intrinsic, a result of multiple successive unresolved low amplitude emission components, or extrinsic to the magnetar, possibly produced by high density plasma clouds traversing the radio beam at high velocities in the pulsar magnetosphere~\citep{Pearlman+2018}.

Multifrequency radio pulse profiles and single pulses from the GC~magnetar revealed that the interstellar scattering is several orders of magnitude smaller than predicted by the NE2001 electron density model~\citep{Cordes+2002, Spitler+2014, Pennucci+2015}. \citet{Spitler+2014} derived a scatter broadening timescale of $\tau_{d}$\,$=$\,1.3\,$\pm$\,0.2\,s at 1\,GHz and a scatter broadening spectral index of \mbox{$\alpha_{d}$\,$=$\,--3.8\,$\pm$\,0.2} from the scattered pulse shapes of single pulses and average pulse profiles between 1.19~and~18.95\,GHz. These results were used to argue for the existence of a single uniform, thin scattering screen at a distance of $\Delta_{\text{GC}}$\,$=$\,5.8\,$\pm$\,0.3\,kpc from the magnetar~\citep{Bower+2014}. Subsequent radio interferometic measurements determined that the angular and temporal broadening are both produced by a single thin scattering screen located $\sim$4.2\,kpc from the magnetar~\citep{Wucknitz2014}. However, \citet{Pearlman+2018} showed that individual single pulses from the GC~magnetar at 8.4\,GHz can be broadened by more than an order of magnitude compared to what is predicted by~\citet{Spitler+2014}, which is incompatible with a static, thin scattering screen at distances\,$\gtrsim$\,1\,kpc. A secondary local screen~(e.g.,~$\sim$0.1\,pc from the magnetar~\citep{Desvignes+2018a}) is not expected to contribute significantly to the temporal broadening~\citep{Dexter+2017}. We refer the interested reader to Sections~4.2 and~4.3 of~\citet{Pearlman+2018} for a detailed discussion of mechanisms which might reconcile the observed pulse broadening behavior.



\section{PSR~J1119--6127: A Transitional Magnetar}
\label{Section:PSR_J1119_6127}

PSR~J1119--6127 is a high magnetic field~($B$\,$\approx$\,4\,$\times$\,10$^{\text{13}}$\,G), rotation-powered pulsar with a spin period of $P$\,$\approx$\,410\,ms~\citep{Majid+2017}. The pulsar resides at the center of the supernova remnant~(SNR)~G292.2--0.5~\citep{Crawford+2001, Pivovaroff+2001, Kumar+2012}, which lies in the Galactic plane at a distance of $\sim$8.4\,kpc~\citep{Caswell+2004}. The pulsar's DM~(707.4\,$\pm$\,1.3\,pc\,cm$^{\text{--3}}$~\citep{He+2013}) and RM~(+853\,$\pm$\,2\,rad\,m$^{\text{2}}$~\citep{Johnston+2006b}) are both large compared to other radio pulsars~(see~Figure~\ref{Figure:Figure2}), and it has one of the largest period derivatives~($\dot{P}$\,$\approx$\,\,4\,$\times$\,10$^{\text{--12}}$~\citep{Antonopoulou+2015}), which implies a characteristic age of $\tau_{c}$\,$\lesssim$\,2\,kyr. PSR~J1119--6127 is the first rotation-powered radio pulsar to display magnetar-like activity. In fact, only one other rotation-powered \text{X-ray} pulsar, PSR~J1846--0258, has previously displayed similar behavior~\citep{Gavriil+2008}, but no radio pulsations have yet been detected from that object. Short \text{X-ray} outbursts were detected from PSR~J1119--6127 on 2016~July~27 and 2016~July~28 with the \textit{Fermi} \text{Gamma-Ray} Burst Monitor~(GBM) and \textit{Swift} Burst Alert Telescope~(BAT), respectively~\citep{Kennea+2016, Younes+2016a}. A large spin-up glitch~\citep{Archibald+2016a, Archibald+2016b}, hardening of the \text{X-ray} spectrum~\citep{Archibald+2016a}, additional \text{X-ray} bursts~\citep{Gogus+2016, Archibald+2018}, and irregular post-outburst timing behavior~\citep{Archibald+2018, Lin+2018} were subsequently reported.

Shortly after the first \text{X-ray} outbursts, we used the DSN's~70\,m~radio telescope, DSS-43, to monitor changes in the pulsar's radio emission during this period of magnetar-like activity. Simultaneous observations were regularly performed at 2.3~and~8.4\,GHz over the course of $\sim$5 months following the initial \text{X-ray} bursts. After this period, the pulsar was intermittently monitored. Several of these observations are described in detail in~\citet{Majid+2017}, and additional results will be presented in an upcoming paper by~\citet{Pearlman+2019b}.

\begin{figure*}[b!]
	\centering
	\includegraphics[trim=0cm 0cm 0cm 0cm, clip=false, scale=0.79, angle=0]{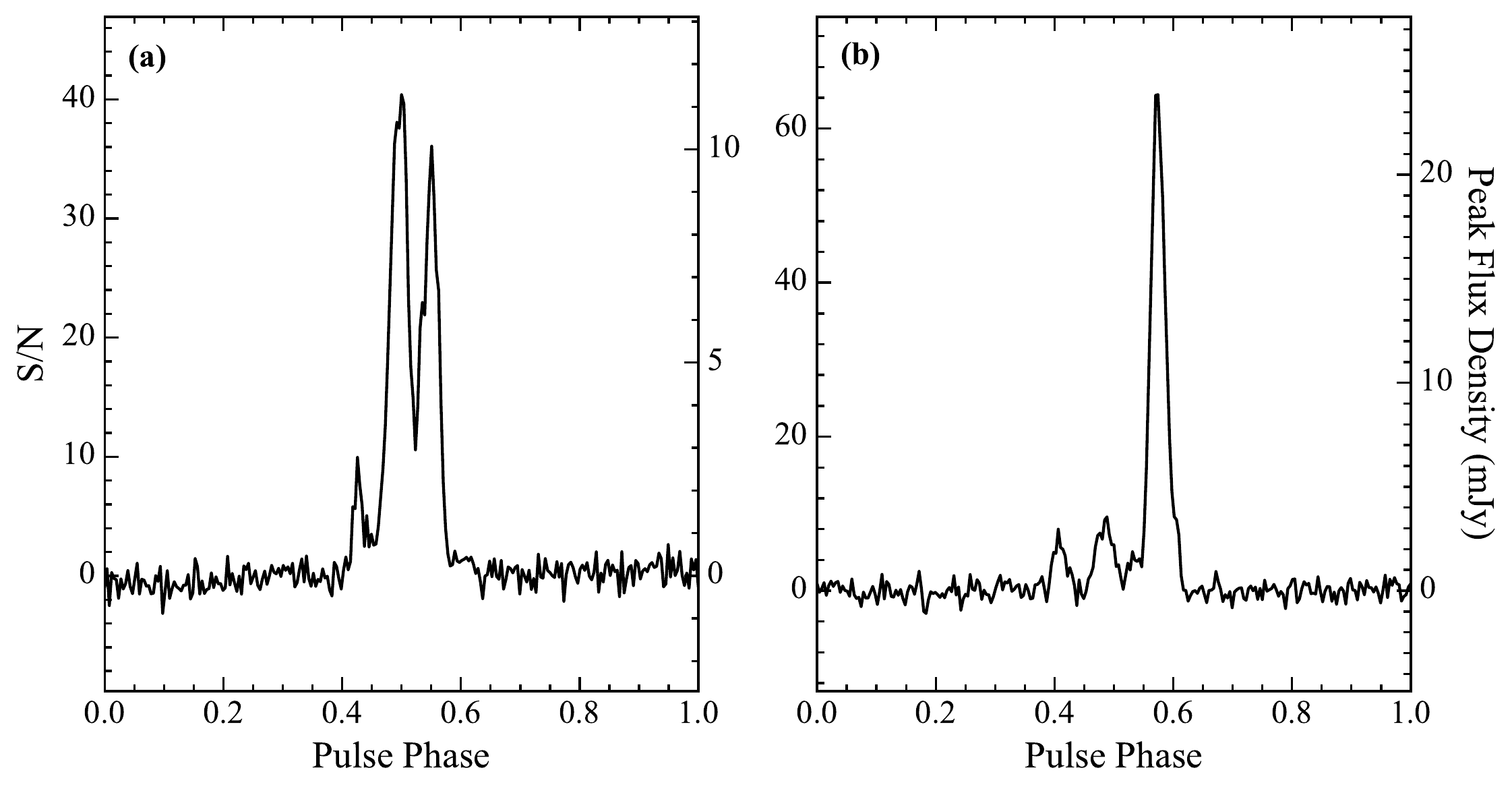}
	\caption{Average $S$-band pulse profiles of PSR~J1119--6127, obtained with the 70\,m DSN radio telescope~(DSS-43) on (a)~2016~August~19 and (b)~2016~September~1, during the pulsar's recent episode of magnetar-like activity. This figure was adapted from~\citet{Majid+2017}. Additional observations during the pulsar's post-outburst recovery period will be presented in~\citet{Pearlman+2019b}.}
	\label{Figure:Figure6}
\end{figure*}

We found that radio pulsations from PSR~J1119--6127 disappeared after the initial \text{X-ray} outbursts, and the emission reactivated approximately two weeks later~\citep{Burgay+2016a, Burgay+2016b, Majid+2016, Majid+2017, Pearlman+2017, Dai+2018, Pearlman+2019b}. The pulse profiles at 2.3~and~8.4\,GHz dramatically evolved over a time period of several months after the radio emission resumed, which is atypical of ordinary radio pulsars~\citep{Majid+2017, Pearlman+2017, Pearlman+2019b}. The 2.3\,GHz pulse profiles developed a multicomponent emission structure, while the 8.4\,GHz profile showed a single emission peak that varied in strength during this time period~\citep{Majid+2017, Pearlman+2017, Pearlman+2019b}. Previous radio observations performed before the bursts indicated that the pulse profile was predominantly single-peaked, and an extremely rare double-peaked structure was seen only once after a strong glitch in~2007~\citep{Camilo+2000, Crawford+2003, Johnston+2006b, Weltevrede+2011}. In contrast, our post-outburst radio observations showed that the 2.3\,GHz pulse profile developed multiple emission components, two of which significantly weakened several weeks after the magnetar-like activity had subsided~(see~Figure~\ref{Figure:Figure6})~\citep{Majid+2017, Pearlman+2017, Pearlman+2019b}.

Simultaneous radio observations at 2.3 and 8.4\,GHz during this period revealed that the spectral index was unusually steep and comparable to the average spectral index of normal radio pulsars when the pulse profiles were multi-peaked. As the profiles stabilized and became single-peaked, the pulsar's radio spectrum flattened considerably~\citep{Pearlman+2016, Majid+2017, Pearlman+2017, Pearlman+2019b}. Significant changes in the pulsar's rotational frequency, flux density, and single pulse behavior were also observed~\citep{Majid+2016, Pearlman+2016, Majid+2017, Pearlman+2017, Pearlman+2019b}. These results demonstrate that PSR~J1119--6127 is a transitional object, i.e.,~a high magnetic field, rotation-powered radio pulsar capable of exhibiting transient magnetar-like behavior.

The detection of pulsed radio emission from several magnetars, along with the behavior observed from PSR~J1119--6127 during its 2016~outburst, has weakened the notion that there is a sharp separation between magnetars and high magnetic field, rotation-powered radio pulsars. Historically, surface dipolar magnetic fields above the quantum critical value ($B_{\text{Q}}$\,$=$\,$m_{e}^2c^3/e\hbar$\,$\approx$\,4.4\,$\times$\,10$^{\text{13}}$\,G) and persistent X-ray luminosities exceeding rotational energy losses ($L_{x}$\,$>$\,$\dot{E}$) were interpreted as some of the defining observational characteristics of magnetars, but these criteria are not always reliable predictors of magnetar-like behavior~\citep{Mereghetti+2008, Rea+2012}. In the case of PSR~J1119--6127, the pulsar's quiescent X-ray luminosity in the 0.5--10\,keV energy band ($L_{x}$\,$=$\,0.9\,$\times$\,10$^{\text{33}}$\,erg\,s$^{\text{--1}}$;~\citealt{Gonzalez+2005}) is several orders of magnitude smaller than its spin-inferred rotational energy ($\dot{E}$\,$=$\,2.3\,$\times$\,10$^{\text{36}}$\,erg\,s$^{\text{--1}}$), i.e.~$L_{x}/\dot{E}$\,$\approx$\,0.0004, which suggests that the pulsar is often primarily powered by its rotation. However, during the 2016~outburst, the pulsar's X-ray luminosity rose to $L_{x}$\,$\sim$\,0.1$\dot{E}$~\citep{Archibald+2016a}. Other radio magnetars, e.g.~PSR~J1622--4950, XTE~J1810--197, and 1E~1547.0--5408, also have X-ray conversion efficiencies below unity ($L_{x}/\dot{E}$\,$<$\,1)~\citep{Rea+2012} and display transient high-energy emission. This suggests that some magnetars and high magnetic field pulsars may be powered through a combination of magnetic and rotational energy.



\section{PSR~J1622--4950}
\label{Section:PSR_J1622_4950}

PSR~J1622--4950 was first detected using the Parkes radio telescope, and it is the only magnetar discovered at radio wavelengths without prior knowledge of a corresponding X-ray counterpart~\citep{Levin+2010}. The pulsar has a spin period of $P$\,$\approx$\,4.326\,s and a DM~of 820\,pc\,cm$^{\text{--3}}$~\citep{Levin+2010}. The timing solution, reported by~\citet{Levin+2010}, implies a very high surface magnetic field of $B$\,$\approx$\,2.8\,$\times$\,10$^{\text{14}}$\,G and a characteristic age of $\tau_{c}$\,$\approx$\,4\,kyr. PSR~J1622--4950 has a flat radio spectrum and a highly variable flux density and pulse profile~\citep{Levin+2010, Keith+2011, Anderson+2012, Levin+2012, Scholz+2017}, which is similar to other radio magnetars.

After its initial discovery, the magnetar was regularly monitored and detected with variable flux densities at the Parkes Observatory until 2014~March~\citep{Scholz+2017}. \citet{Scholz+2017} resumed monitoring the magnetar in 2015~January, but they failed to detect the pulsar through 2016~September. After remaining in a dormant state for roughly 2~years, PSR~J1622--4950 resumed its radio emission sometime between January and April of~2017~\citep{Camilo+2018}. Shortly after the radio reactivation, we carried out simultaneous observations of PSR~J1622--4950 at $S$-band and $X$-band using the 70\,m radio telescope, DSS-43, on 2017~May~23~\citep{Pearlman+2017b}. \citet{Pearlman+2017b} found that the magnetar's radio spectrum had significantly steepened between 2.3~and~8.4\,GHz, and its spectral behavior was consistent with the majority of ordinary pulsars.

Beginning in 2017~April, we used the 34\,m~DSN~radio dishes near Canberra, Australia to initiate a monitoring program to observe PSR~J1622--4950 over $\sim$30 epochs to date, already lasting more than a year. The observations were made using simultaneous $S$/$X$-band receivers. The data processing steps and observational setup were described in detail by~\citet{Majid+2017}. The details of this monitoring campaign will be presented in~\citet{Pearlman+2019a}. The observing epochs were not regularly spaced during the year-long monitoring campaign due to various logistical issues, including scheduling constraints. Each observing epoch ranged in duration from roughly half an hour to four hours in length. The duration of each epoch was sufficient to obtain accurate measurements of the flux densities in both observing bands, spectral index, pulsar spin period, rate of single pulse emission, and pulse scatter broadening.

\begin{figure*}[b!]
	\centering
	\includegraphics[trim=0cm 0cm 0cm 0cm, clip=false, scale=0.78, angle=0]{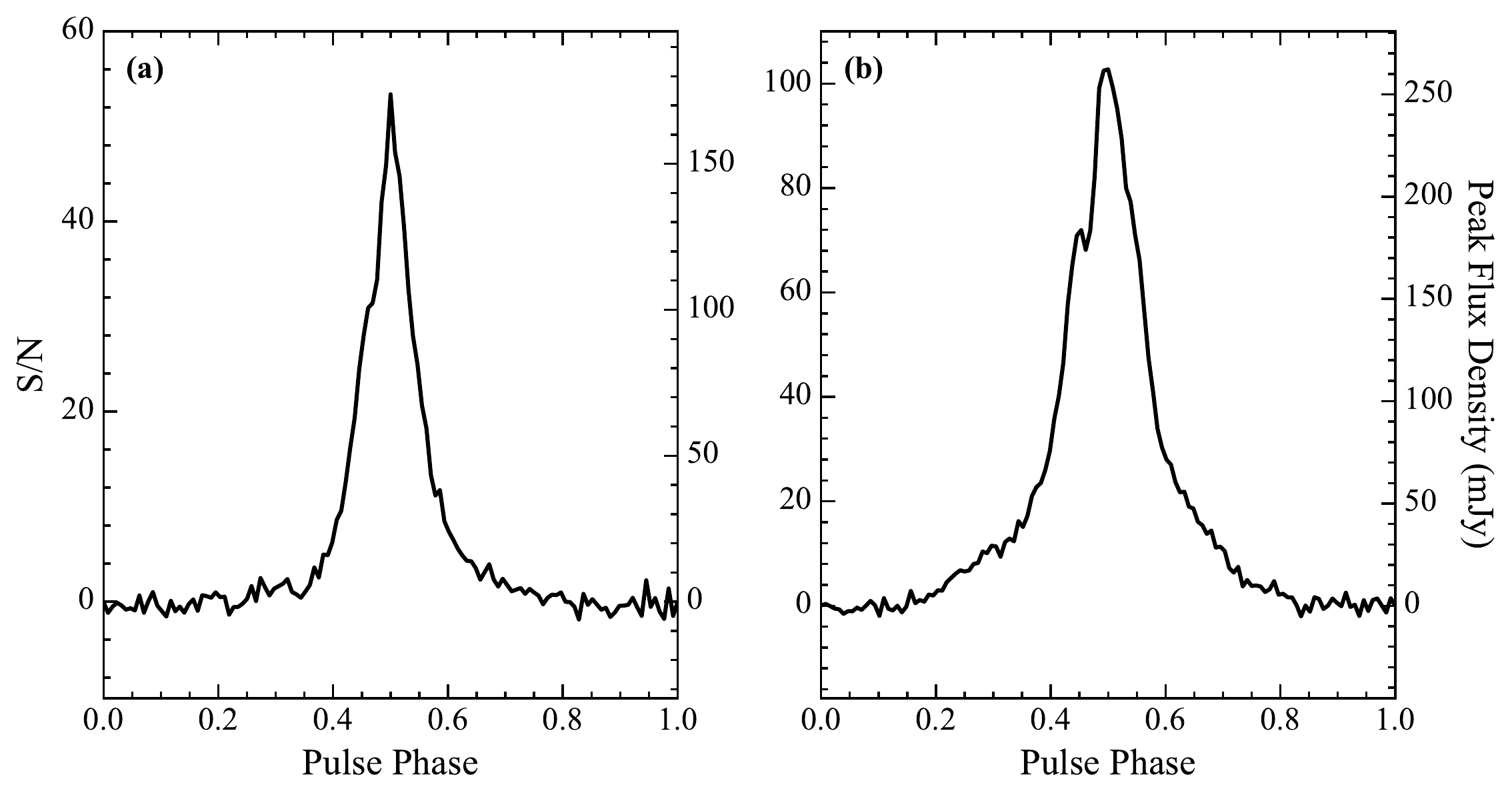}
	\caption{Average $S$-band pulse profiles of the radio magnetar, PSR~J1622--4950, obtained using the 34\,m radio telescopes near Canberra, Australia on (a)~2018~April~26 and (b)~2018~May~10. Additional details on these observations will be presented in~\citet{Pearlman+2019a}.}
	\label{Figure:Figure7}
\end{figure*}

\begin{figure*}[b!]
	\centering
	\includegraphics[trim=0cm 0cm 0cm 0cm, clip=false, scale=0.6, angle=0]{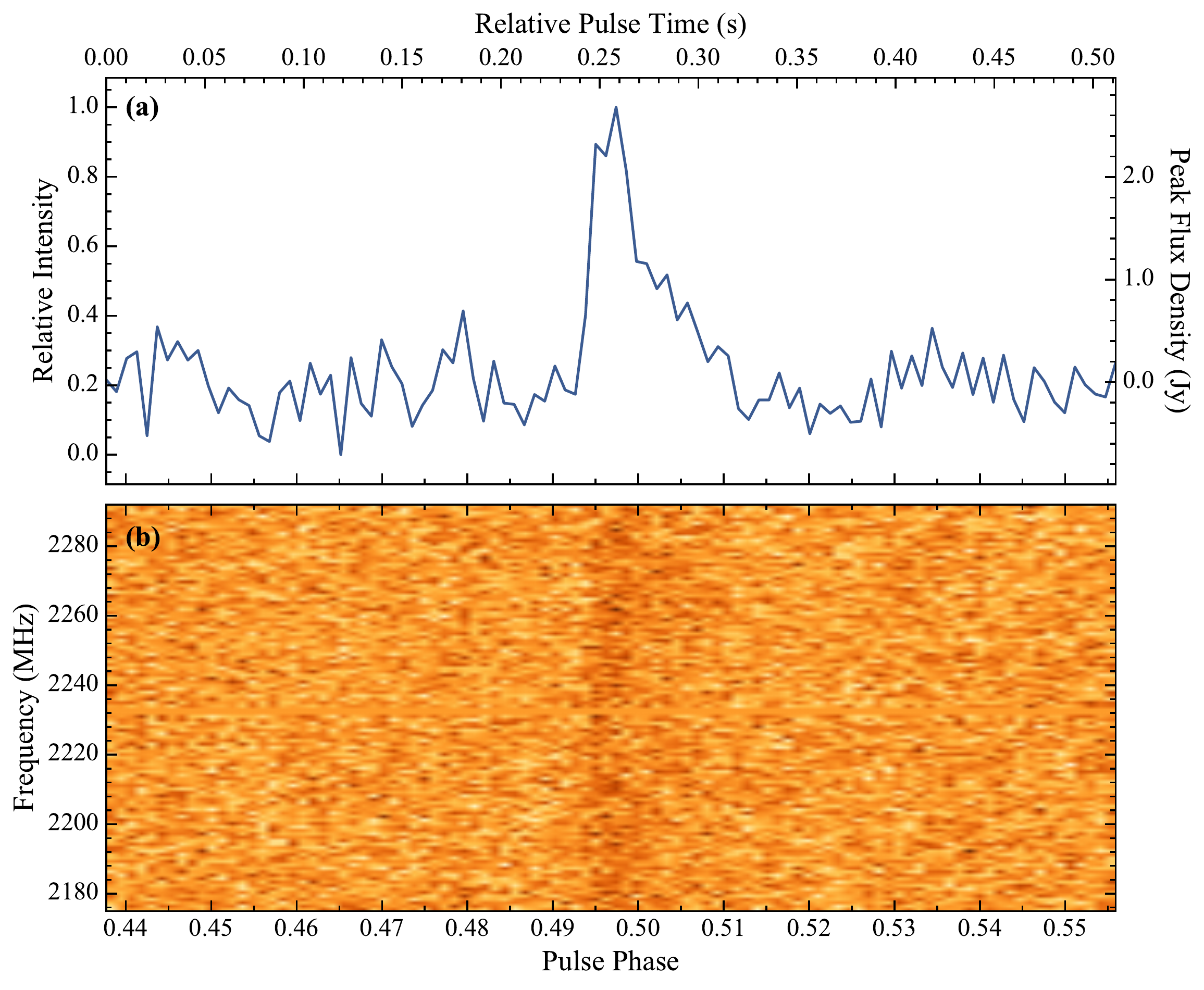}
	\caption{Example of a bright single pulse detected at $S$-band from PSR~J1622--4950 during the observation on 2018~May~10~(see Figure~\ref{Figure:Figure7}(b)). In panel (a), we show the integrated single pulse profile with a time resolution of $\sim$5\,ms. The dynamic spectrum, dedispersed at the magnetar's nominal~DM of 820\,pc\,cm$^{\text{--3}}$, is shown in panel (b). This event has a S/N~of~$\sim$10 and was detected by convolving the dedispersed time-series with a $\sim$23\,ms wide boxcar template. The shape of the single pulse is noticeably scatter-broadened.}
	\label{Figure:Figure8}
\end{figure*}

Our results indicate that the magnetar exhibited noticeable changes in its pulse profile at $S$-band~(e.g.,~see~Figure~\ref{Figure:Figure7}), but also particularly at $X$-band. PSR~J1622--4950's pulse profiles sometimes showed evidence of the emergence of a new pulse component. The magnetar's flux density also showed variability in the range of a few~mJy to tens of~mJy. In addition, we also observed remarkable short-term changes in the magnetar's emission behavior during a few observing epochs and detected bright single pulses with varying pulse morphology~(e.g.,~see~Figure~\ref{Figure:Figure8}).



\section{XTE~J1810--197}
\label{Section:XTE_J1810_197}

In July~2003, XTE~J1810--197 was serendipitiously discovered using the \textit{Rossi X-ray Timing Explorer}~(\textit{RXTE}) after the magnetar underwent a transient X-ray outburst~\citep{Ibrahim+2004}. The pulsar's rotational period was determined to be $P$\,$\approx$\,5.54\,s, and a high spin-down rate of $\dot{P}$\,$\approx$\,10$^{\text{--11}}$ was measured, which suggests that the neutron star is young~($\tau$\,$\approx$\,7.6\,kyr) and possesses a large spin-inferred dipolar magnetic field~($B$\,$\approx$\,2.6\,$\times$\,10$^{\text{14}}$\,G). Approximately one year later, a coincident radio source was found at the location of the pulsar~\citep{Halpern+2005}. Subsequently, pulsed radio emission was detected from the magnetar, making XTE~J1810--197 the first magnetar with detected radio pulsations~\citep{Camilo+2006}. Multifrequency radio observations between 0.7~and~42\,GHz revealed that the magnetar emits bright, highly linearly polarized radio pulses, comprised of narrow sub-pulses with widths $\le$\,10\,ms, during each rotation~\citep{Camilo+2006}. These results demonstrated that there is an underlying connection between magnetars and the larger population of ordinary radio pulsars.

Radio pulsations from XTE~J1810--197 suddenly ceased without warning in late~2008, despite continued X-ray activity~\citep{Camilo+2016, Pintore+2016, Pintore+2019}. After more than 10 years in quiescence, bright radio pulsations were detected again on 2018~December~8 with the 76\,m Lovell Telescope at Jodrell Bank~\citep{Lyne+2018}. Since the magnetar's reactivation, numerous radio observatories have carried out follow-up observations of the magnetar~\citep{Desvignes+2018b, Lower+2018, Joshi+2018, DelPalacio+2018, Majid+2019, Trushkin+2019}. In particular, we observed XTE~J1810--197 continuously for 5.5 hours on 2018~December~25~(MJD~58477.05623) using one of the DSN's 34\,m radio telescopes near Canberra, Australia~\citep{Majid+2019}. Right circular polarization data were simultaneously recorded at center frequencies of 8.4~and~32\,GHz, with roughly 500\,MHz of bandwidth at each frequency band, using the JPL~ultra-wideband pulsar machine. Our best estimate of the barycentric spin period and DM is 5.5414471(5)\,s and 178\,$\pm$\,9\,pc\,cm$^{\text{--3}}$, respectively. The average pulse profiles were noticeably variable, and we detected bright, multi-component single pulses at both frequency bands. We measured mean flux densities of 4.0\,$\pm$\,0.8\,mJy at 8.4\,GHz and 1.7\,$\pm$\,0.3\,mJy at 32\,GHz, which yielded a spectral index of --0.7\,$\pm$\,0.2 over this wide frequency range~\citep{Majid+2019}. Additional multifrequency observations are needed to study the magnetar's behavior after its recent outburst. Towards this end, we are continuing to carry out high frequency radio observations of XTE~J1810--197 using the DSN's 70 and 34\,m radio dishes near Canberra. A regular monitoring program of this magnetar is also planned.


\section{Discussion and Conclusions}
\label{Section:Conclusions}

We have presented an overview of recent results from observations of the radio magnetars, PSR~J1745--2900, PSR~J1622--4950, and XTE~J1810--197, and the transitional magnetar candidate, PSR~J1119--6127, obtained using the DSN~radio telescopes near Canberra, Australia. These studies provide further evidence of the variable nature of these objects. Each of these radio magnetars exhibited remarkable pulse profile changes over timescales of weeks to months, with large accompanied flux and spectral index variations. In combination with results at X-ray wavelengths, the variability observed in radio magnetars may be explained by the conditions of the magnetosphere~\citep{Turolla+2015, Archibald+2017, Kaspi+2017}, though the level of variability in each object likely depends on the size and geometry of magnetospheric deformations. Toroidal oscillations in the star may be excited during an outburst, which then modify the magnetospheric structure and allow radio emission to be produced. Since this emission behavior is clearly transitory, further radio monitoring of these objects is needed to study their long-term radiative and timing behavior.

The~DSN has served as an excellent facility for performing \text{state-of-the-art} pulsar observations, as we have demonstrated through the study of the four magnetars discussed in this paper. The combination of the excellent sensitivity of the DSN~antennas, particularly with the presence of a large 70\,m diameter dish at each of the DSN~complexes, multifrequency receivers, and the recent deployment of modern pulsar machines, offers an opportunity for pulsar observations that will be a significant addition to the already existing resources in pulsar astronomy. The availability of the 70\,m antenna in Canberra, with its southern location, makes it an ideal resource, complementing the Parkes telescope, for observations of Galactic plane sources, including the Galactic Center. In search mode, the DSN's pulsar machines offer high frequency and timing resolution with the ability to record multiple frequencies and incoming polarization bands simultaneously. This allows for observations with a high instantaneous sensitivity, which are quite useful for studies of single pulses. With precision tracking capabilities available at multiple frequencies, the~DSN is particularly \text{well-suited} for carrying out observations at shorter wavelengths, which have proven useful for studying objects such as magnetars with flatter spectral indices and high~DM pulsars.


\section{Acknowledgments}

A.~B.~Pearlman acknowledges support by the Department of Defense~(DoD) through the National Defense Science and Engineering Graduate~(NDSEG) Fellowship Program and by the National Science Foundation~(NSF) Graduate Research Fellowship under Grant~No.~\text{DGE-1144469}.

We thank the Jet Propulsion Laboratory's Research and Technology Development program and Caltech's President's and Director's Fund for partial support at~JPL and the Caltech~campus. A portion of this research was performed at the Jet~Propulsion~Laboratory, California~Institute~of~Technology and the Caltech campus, under a Research and Technology Development Grant through a contract with the National Aeronautics and Space Administration. U.S.~government sponsorship is acknowledged.

\clearpage

\bibliography{references}

\end{document}